\documentclass{article}

\font\sqi=cmssq8
\def\beq{\begin{equation}}
\def\eeq{\end{equation}}
\def\DR{\rm I\kern-1.45pt\rm R}
\def\DC{\kern2pt {\hbox{\sqi I}}\kern-4.2pt\rm C}
\def\DH{\rm I\kern-1.5pt\rm H\kern-1.5pt\rm I}

\newcommand{\cH}{{\cal H}}
\renewcommand{\imath}{{\rm i}}
\begin{document}
\begin{center}
{\Large\bf Non-Commutative Corrections
to the  MIC-Kepler Hamiltonian}\\
\vspace{0.5 cm}
{\large\bf Dennis  Khetselius}\\
\vspace{0.5 cm}
{\it Department of Physics and Astronomy, University of New Mexico, Albuquerque, NM 87131} \\
dkhets@yahoo.com
\end{center}
\begin{abstract}
Non-commutative corrections to the MIC-Kepler System (i.e. hydrogen atom
 in the presence of a magnetic monopole) are computed in Cartesian and
parabolic coordinates. Despite the fact that there is no simple analytic
expression for non-commutative perturbative corrections to the
MIC-Kepler spectrum, there is
a term that gives rise to the linear Stark effect which didn't exist in the
standard hydrogen model.
\end{abstract}
\vspace{1cm}
Progress in the String theory has recently ignited a great deal of
interest to field theories in non-commutative spaces \cite{witten}.
Quantum Hall effect \cite{Polychronakos:2001mi}, non-commutative classical and quantum mechanics
\cite{sj}- \cite{Bellucci:2002yh}, and various non-commutative
phenomenological models \cite{Mazumdar:2001jc} are among these types of theories. In addition to
the string theory, these models also give rise to
theories describing particles with spin.
While most of the research was focused on
mathematical aspects of two-dimensional quantum mechanics, three-dimensional quantum mechanics also attracted some attention.
The later was inspired by possible phenomenological consequences of
the non-commutativity in three dimensions.
In their pioneer paper Chaichian et al. \cite{sj} have found non-commutative
corrections to the hydrogen atom spectrum and discovered a new channel induced by these corrections to the Lamb shift. Specific form
of the vertex in non-commutative quantum field theory (NCQED) resulted in lack
of similar corrections to the Stark effect.
Here we consider a non-commutative MIC-Kepler model - an elegant generalization of the Coulomb system. Integrable MIC-Kepler system, originally constructed by Zwanziger
\cite{Z} and later rediscovered by McIntosh and Cisneros \cite{mic} is characterized by the presence of an external monopole field.
Such a system is described by the Hamiltonian
\begin{equation}
{\cal H}_0=\frac{\hbar^2}{2\mu}(i{\bf{\nabla}}+ {s}{\bf A})^2
+\frac{\hbar^2{s}^2}{2\mu r^2}-\frac{\gamma}{r},\quad{\rm
where}\quad {\rm rot}{\bf A}=\frac{{\bf r}}{r^3}. \label{1}
\end{equation}
Its distinctive peculiarity lies in the hidden symmetry
given by the following constants of motion 
\begin{equation}
{\bf I}=\frac{\hbar}{2\mu}\left[(i{\bf\nabla}+ {s}{\bf
A})\times{\bf J}-{\bf J}\times (i{\bf\nabla}+ {s}{\bf A})\right]
+\gamma\frac{{\bf r}}{r},\quad {\bf J}=-\hbar (i{\bf\nabla}+
{s}{\bf A})\times{\bf r}
 +\frac{\hbar{s}{\bf r}}{r}.
\label{2}
\end{equation}
These constants of motion together with the Hamiltonian, form the
algebra of quadratic symmetry of the Coulomb model.
Operators ${\bf J}$ and ${\bf I}$ define
the angular momentum and the Runge-Lenz vector for the MIC-Kepler system.
For fixed negative values of energy, these constants of motion
form $SO(4)$ algebra and for positive values of energy -
$SO(3.1)$ one. Due to the hidden symmetry, the MIC-Kepler problem
could be factorized in some coordinate systems, e.g. in spherical and parabolic coordinates.
The MIC-Kepler system, therefore, becomes a natural generalization of the
Coulomb model in the presence of Dirac's monopole.
Monopole number $s$ satisfies the Dirac's charge quantization rule,
 $s=0,\pm1/2,\pm 1,\ldots$.

The MIC-Kepler system inherits most of the properties of hydrogen atom.
Its behavior however, becomes qualitatively different in a static electric field.
In the case of the MIC-Kepler model, degeneracy in the energy spectrum
is removed by a constant electric field. In contrast with
the usual hydrogen atom introduction of such a field gives
rise to non-trivial corrections to the linear Stark effect \cite{mn}. Therefore, one can expect that the non-commutative corrections yield qualitative difference between
the MIC-Kepler and hydrogen atom, at least in the context of Stark effect.\\

Our discussion we begin with a description of the MIC-Kepler system
in the framework of non-commutative geometry and then, we calculate the
corresponding non-commutative corrections. Non-commutativity in Cartesian coordinates
can be introduced via the following non-commutative addition to the Poisson brackets:
\begin{equation}\label{nc}
[x_i,x_j]=i\theta_{ij},
\end{equation}
where $\theta_{ij}$ are $c$-numbers of dimensionality
(length)$^{2}$.\\ \noindent
We should point out that in non-commutative geometry ordinary
product of fields is replaced by the so-called Weyl-Moyal "star" product:
\begin{equation}
(f*h)(x) = exp(\frac{i}{2}\theta_{\mu \nu} \frac{\partial}{\partial x_{\mu}}\frac{\partial}{\partial y_{\nu}})f(x)h(y)|_{x=y}
\end{equation}
where $f$ and $h$ are infinitely differentiable functions.
One can show \cite{mez} that in the case where $[p_i,p_j]=0$, non-commutative quantum mechanics is
reduced to the usual QM described by the Schr\"odinger equation
\begin{equation}
\cH (p,{\hat x})\Psi({\hat x})=E \Psi({\hat x}), \mbox{ where } {\hat x}_i=x_i+\frac{1}{2} \theta_{ij}p_j.
\end{equation}
The non-commutative Hamiltonian indeed turns out to be equivalent to a commutative one
and the difference between non-commutative and ordinary quantum mechanics simply
becomes a choice of polarization.
\beq {\hat{\cal
 H}}=\frac{{\bf p}^2}{2\mu}+ V({\bf x}), \label{h0}
 \eeq
Presence of a magnetic field makes the description
 of a non-commutative system much more involving both technically and conceptually \cite{Mazumdar:2001jc}, \cite{np}.
Indeed, formal addition of the nontrivial commutator (1) to our system violates
the Jacobi identities. This is hardly surprising: the magnetic field itself becomes
non-commutative and does no longer correspond to a closed two-form.
The correct way would be 
first to replace the vector-potential of a commutative 
magnetic field with the vector potential of a non-commutative one 
and then, to replace the non-commutative Schr\"odinger equation with the equivalent
commutative one. In the context of non-commutative mechanics this
scheme was originally discussed in \cite{hor}.
This transformation (known as a Seiberg-Witten map) was found for the case of
constant magnetic field (e.g.  Mezinchescu \cite{mez}). Our aim here is to construct similar  mapping for an inhomogeneous field of a magnetic monopole.\\
\vspace{0.3cm}
  
Classical commutative MIC-Kepler system is described by the Hamiltonian
\begin{equation}
H=\frac{\mathbf{p}^2}{2} + \frac{s^2}{2\mathbf{x}^2} - \frac{\gamma}{x}
\end{equation}
together with the symplectic structure
\begin{equation}
\Omega_0 = \mathbf{dp\wedge dx} + s \frac{\epsilon_{ijk}x_idx_j \wedge dx_k}{x^3}
\end{equation}
On the phase space $(\mathbf{x},\mathbf{p})$
 this structure determines a set of non-canonical Poisson brackets
\begin{eqnarray}
[p_i,p_j] &=& s \frac{i \epsilon_{ijk}x_k}{x^3}  \\ \nonumber
[p_i,x_j] &=& -i \delta_{ij}
\end{eqnarray}
where we use the convention $\hbar=1$. \\ \noindent
In order to include non-commutative corrections one should add the corresponding term
to the original symplectic form
\begin{eqnarray}
\Omega &=& \Omega_0 + \frac{\theta_i}{2} \epsilon_{ijk} dp_j \wedge dp_k, \label{o1} \\
\theta_i &=& \epsilon_{ijk}\theta_{jk} \nonumber
\end{eqnarray}
where $\mathbf{\theta}$ is a perturbation parameter.\\ \noindent
Our goal is to find such mapping
\begin{equation}
\mathbf{p} \rightarrow \mathbf{\tilde{p}}, \mathbf{x} \rightarrow \mathbf{\tilde{x}}
\end{equation}
that transforms our Poisson brackets into the canonical form
\begin{eqnarray}
[\tilde{p}_i,\tilde{x}_j] = -i \delta_{ij} + O(\theta^2)   \\ \nonumber
[\tilde{x}_i,\tilde{x}_j] = [\tilde{p}_i,\tilde{p}_j] = 0 + O(\theta^2)
\end{eqnarray}
The first step is to eliminate the second, non-canonical term in the symplectic 
structure (2). To do this, in Cartesian coordinates (see e.g. Mardoyan et al \cite{mn})
 we introduce the Dirac's monopole potential
$A_i = \frac{\epsilon_{3ij}x_j}{x(x-x_3)}$ together with the mapping
$\mathbf{\hat{p}} = \mathbf{p} + s\mathbf{A}, \mathbf{\hat{x}} = \mathbf{x}$.
It it straightforward to verify that this mapping indeed serves our purpose 
to the zeroth power in the
parameter $\theta$.
\begin{equation}
\Omega = \mathbf{d\hat{p} \wedge d\hat{x}} + \frac{\theta_i}{2} \epsilon_{ijk} d\hat{p}_j \wedge d\hat{p}_k + \frac{s^2 \theta_i}{2} \epsilon_{ijk}dA_j \wedge dA_k - s \theta_i \epsilon_{ijk}d\hat{p}_j \wedge dA_k
\end{equation}
In order to complete our task in the first order, we introduce another
set of transformations
\begin{eqnarray}
\tilde{p}_{i} &=& \hat{p}_{i} + \frac{s^2 \theta_{3} \epsilon_{3ij} \hat{x}_{j}}{4\hat{x}^2(\hat{x}-\hat{x}_3)^2} \\ \nonumber
\tilde{x}_{i} &=& \hat{x}_{i} + \frac{\theta_j}{2}\epsilon_{ijk}(\hat{p}_{k} - 2sA_{k})
\end{eqnarray}
so that the symplectic structure in the new coordinates takes the canonical form
\begin{equation}
\Omega = \mathbf{d\tilde{p} \wedge d\tilde{x}} + O(\theta^2)
\end{equation}
The original field Hamiltonian becomes dependent on the non-commutativity parameter $\theta$. Reverting the coordinate basis $p, x \rightarrow \tilde{p}, \tilde{x}$ we deduce the corresponding linear corrections:
\begin{eqnarray}
H &=& \frac{(\mathbf{\tilde{p}} - s\mathbf{A(\tilde{x})})^2}{2} + \frac{s^2}{2(\tilde{x}_i + \frac{1}{2}\epsilon_{ijk}\theta_j(\tilde{p}_k - 2sA(\tilde{x})_k))^2}  - \\
\nonumber & &\frac{\gamma}{\tilde{x}_i + \frac{1}{2}\epsilon_{ijk}\theta_j(\tilde{p}_k - 2sA(\tilde{x})_k)}
 - \frac{s^2\theta_i \tilde{x}_i \epsilon_{3jk}\tilde{x}_j(\tilde{p}_k - sA(\tilde{x})_k) }{4\mathbf{\tilde{x}}^2(\tilde{x} - \tilde{x}_3)^2}
\end{eqnarray}
\begin{equation}
H_{corr} = \frac{s^2 (\mathbf {\theta * \tilde{x}}) (\mathbf{A(\tilde{x})*(\tilde{p}- sA(\tilde{x})))}}{4\mathbf{\tilde{x}}(\tilde{x} - \tilde{x}_3)} -
 \frac{\mathbf{\theta * \tilde{x} \times (\tilde{p} - 2sA)}}{2\tilde{x}^3}
(\frac{s^2}{\tilde{x}} - \gamma) + O(\theta^2)
\end{equation}
To find the energy spectrum we rewrite the last expression in parabolic coordinates:
\begin{eqnarray}
\tilde{x}_1 &=&  \sqrt{\zeta \eta}cos(\phi) \\ \nonumber
\tilde{x}_2 &=&  \sqrt{\zeta \eta}sin(\phi) \\ \nonumber
\tilde{x}_3 &=& \frac{1}{2}(\zeta - \eta)
\end{eqnarray}
\begin{eqnarray}
& &H_{corr} = (\frac{4\gamma}{(\zeta + \eta)^3} - \frac{8s^2}{(\zeta + \eta)^4})
(\theta_{1}(\frac{2\sqrt{\zeta \eta}sin(\phi)}{\zeta + \eta}(\zeta \frac{\partial}
{\partial \zeta} - \eta \frac{\partial}{\partial \eta}) - \\ \nonumber
& &2\frac{\sqrt{\zeta \eta}(\zeta - \eta)sin(\phi)}{\zeta + \eta}(\frac{\partial}
{\partial \zeta} + \frac{\partial}{\partial \eta}) - \frac{(\zeta - \eta)cos(\phi)}
{2\sqrt{\zeta \eta}}\frac{\partial}{\partial \phi} + 2s\sqrt{\frac{\zeta}{\eta}}
\frac{\zeta - \eta}{\zeta + \eta}cos(\phi)) + \\ \nonumber
& &\theta_{2}(\frac{2\sqrt{\zeta \eta}cos(\phi)}{\zeta + \eta}(\zeta \frac{\partial}
{\partial \zeta} - \eta \frac{\partial}{\partial \eta}) - 2\frac{\sqrt{\zeta \eta}
(\zeta - \eta)cos(\phi)}{\zeta + \eta}(\frac{\partial}{\partial \zeta} + \frac{\partial}
{\partial \eta}) - \\ \nonumber & &\frac{(\zeta - \eta)sin(\phi)}{2\sqrt{\zeta \eta}}
\frac{\partial}{\partial \phi} + 2s\sqrt{\frac{\zeta}{\eta}}\frac{\zeta - \eta}
{\zeta + \eta}sin(\phi)) +
\theta_3(\frac{\partial}{\partial \phi} - 4s\frac{\zeta}{\zeta+\eta})) + \\ \nonumber
& &\frac{s^2}{x^2(x-x_3)^2}(\theta_1 \sqrt{\eta \zeta}cos(\phi) + \theta_2\sqrt{\eta
\zeta}sin(\phi)  + \theta_3 \frac{\zeta - \eta}{2 \zeta \eta (\zeta + \eta)^2})(\frac{\partial}{\partial \phi} - 2s\frac{\zeta}{\zeta+\eta})
 \label{sh}
\end{eqnarray}
To deal with such a complex Hamiltonian we make the ansatz $\theta_{3} = \theta, \theta_{2} = \theta_{1} = 0$. Then, the linear corrections become
\begin{equation}
H_{corr} = \theta((\frac{4\gamma}{(\zeta + \eta)^3} - \frac{8s^2}{(\zeta + \eta)^4})
(\frac{\partial}{\partial \phi} - 4s\frac{\zeta}{\zeta+\eta})\\ \nonumber
+ \frac{s^2(\zeta - \eta)}{2\eta^2 (\zeta + \eta)^2}(\frac{\partial}{\partial \phi} -
2s\frac{\zeta}{\zeta+\eta}))
\end{equation}
Neither a powerful technique of separation of variables,
nor a further simplifying assumption of an incident s-wave \cite{11}, \cite{12}
\begin{eqnarray}
F(\zeta,\eta,\phi) &=& \Phi(\zeta) \Psi (\eta) e^{im\phi} \nonumber \\
H_{corr}^{m=0} &=& -\theta\frac{\zeta}{(\zeta + \eta)^4}(16s\gamma
 - \frac{32s^3}{(\zeta + \eta)^2} + \frac{s^3(\zeta - \eta)}{\eta^2})
\end{eqnarray}
enable us to solve this equation (here we defined $F(\zeta,\eta,\phi)$ as the monopole wave function).
Despite the difficulty to find explicit form of the linear corrections to the spectrum, some conclusions still could be drawn from our
discussion. According to the arguments presented by Chaichian et al \cite{sj}, the  absence of linear corrections to the Stark effect is a
consequence of a specific form of the electron-photon vertex function at tree level in NCQED.
\begin{eqnarray}
& & \Gamma_i = exp(\frac{i}{2} \mathbf{p \times p'}) \gamma_i \\
& & \mathbf{p \times p'} = p_i \theta_{ij} p_{j}'
\end{eqnarray}
(we still use the convention $\hbar=1$).\\ \noindent
The second term in the expansion in $\theta$ gives rise to a non-commutative correction to the electric dipole momentum
\begin{equation}
<P_i>_{\theta} = \frac{1}{2} e \theta_{ij} p_j
\end{equation}
Presence of the magnetic monopole field modifies the last expression:
\begin{equation}
<P_i>_{\theta} = \frac{1}{2} e \theta_{ij} \tilde{p}_j - \frac{se}{2} \theta{ij}A(\tilde{x})_j + O(\theta^2)
\end{equation}
Just as in the commutative case, the first term gives no contribution to linear corrections to the Stark effect.
The second term however, results in non-trivial corrections to the energy spectrum - the linear non-commutative Stark effect
\begin{equation}
\Delta E_{Stark}^{\theta} = \frac{s}{2}\mathbf{\theta} \times \mathbf{\rm{E}}<n|\mathbf{A}|n'>
\end{equation}
where n is a set of quantum numbers describing an MIC-Kepler state.\\
\vspace{0.5cm}

Let us briefly recapture some basic results presented above.
We constructed a consistent non-commutative deformation of
the MIC-Kepler system (hydrogen atom in the magnetic field of Dirac's monopole) and computed the first order corrections to its Hamiltonian. It turned out that
the corresponding Schr\"odinger equation is much more complicated than the one for  hydrogen atom.
 Due to the hidden symmetry, the Stark effect spectrum could be easily
 calculated both for the case of hydrogen atom and for the
 MIC-Kepler system\cite{12,mn}. While non-commutativity
does not lead to any additional corrections to the Stark effect
in hydrogen atom \cite{sj}, its contribution to the Stark
effect in the MIC-Kepler system is non-trivial. The task of obtaining a solution to the Schr\"odinger
equation (\ref{sh}) (perhaps with some other simplifying assumptions) remains a subject for future research. Once a solution is found, one will be able to
calculate these corrections explicitly.
\vspace{1cm}

Author would like to thank Armen Nersessian for his contribution to this paper. His
original ideas and numerous helpful comments have made the completion of this work possible.

\end{document}